\documentclass{article}

\usepackage{microtype}
\usepackage{graphicx}
\usepackage{subfigure}
\usepackage{booktabs} 
\usepackage{comment}

\usepackage{hyperref}

\usepackage[accepted]{tccmlicml2021}

\icmltitlerunning{Climate risk disclosure in annual reports}

\begin{document}

\twocolumn[
\icmltitle{Automated Identification of Climate Risk Disclosures in Annual Corporate Reports}




\begin{icmlauthorlist}
\icmlauthor{David Friederich}{bern}
\icmlauthor{Lynn H.~Kaack}{dgess,istp}
\icmlauthor{Alexandra Luccioni}{mila}
\icmlauthor{Bjarne Steffen}{dgess,istp}
\end{icmlauthorlist}

\icmlaffiliation{bern}{Department of Economics, University of Bern, Bern, Switzerland}
\icmlaffiliation{dgess}{Department of Humanities, Social and Political Sciences, ETH Zurich, Zurich, Switzerland}
\icmlaffiliation{istp}{Institute of Science, Technology, and Policy, ETH Zurich, Zurich, Switzerland}
\icmlaffiliation{mila}{Mila -- Quebec AI Institute, Montreal, Canada}

\icmlcorrespondingauthor{Bjarne Steffen}{bjarne.steffen@gess.ethz.ch}
\icmlcorrespondingauthor{Lynn Kaack}{lynn.kaack@gess.ethz.ch}

\icmlkeywords{Machine Learning, ICML}

\vskip 0.3in
]



\printAffiliationsAndNotice{} 

\begin{abstract}
It is important for policymakers to understand which financial policies are effective in increasing climate risk disclosure in corporate reporting. 
We use machine learning to automatically identify disclosures of five different types of climate-related risks. 
For this purpose, we have created a dataset of over 120 manually-annotated annual reports by European firms.
Applying our approach to reporting of 337 firms over the last 20 years, we find that risk disclosure is increasing. Disclosure of transition risks grows more dynamically than physical risks, and there are marked differences across industries. Country-specific dynamics indicate that regulatory environments potentially have an important role to play for increasing disclosure.
\end{abstract}

\section{Introduction}

Climate-related financial disclosures aim at increasing the transparency to guide investment and lending decisions in the financial sector.
They reflect not only the physical effects of a changing climate, also the transition processes to decarbonize the economy that bear financial risks and opportunities. 
Such transition risks and opportunities include policy and legal change, technology and market shifts, and reputation-related risks \citep{climatewise2019transition, doi:10.1002/wcc.678}. 
Transition risks are found to be more imminent than physical risks from climate change, and there are indications that some of those risks are already financially priced \citep{kolbel2020does}. In fact, the G20 Financial Stability Board has established a Task Force on Climate-related Financial Disclosures (TCFD) to issue recommendations for disclosure of climate-related risks and opportunities (CRO) in corporate reporting \citep{tcfd2019}.  
However, recent analyses of the current standard of reporting have found insufficient clarity of CRO reporting (meaning that disclosures are not specific enough to allow a judgement whether CRO are material, i.e.~significant, for future financial performance), with some recent improvements \citep{tcfd2019, cdsb2020falling, demaria2019}. 
To establish more stringent reporting standards on CRO, policy makers and financial regulators have begun establishing regulations \citep{steffen2021comparative}, for example the 2014 {\it EU Non-Financial Reporting Directive} (EU directive 2014/95/EU) and subsequent guidelines (2017/C 215/01 and 2019/C 209/01) \citep{EU2021law},
Art.~173 of the French Energy Transitions law from 2015 \citep{france2015law}, and a recent Executive Order on Climate-Related Financial Risk in the US \citep{EO2021US}.

To monitor the state of climate-related disclosure and establish the effectiveness of related regulation, financial reports need to be analyzed to assess the extent and quality of such disclosures by companies and organizations.
Typically, these analyses are conducted manually, which is time-intensive: recent examples covered only the 40 
\citep{demaria2019} and 50 largest companies \citep{cdsb2020falling} of different listings. Constraining the analysis to only a small set of the largest companies risks introducing bias or preventing important insights: first, the financially strongest companies might not sufficiently represent sectors that are most relevant for low-carbon transitions (e.g., most freight shipping companies are typically very small, while having a significant carbon footprint~\citep{teter2017future}). Second, there are equity concerns if CROs of smaller companies are not monitored (e.g., they could experience significant transition-related opportunities but investments are not directed to them). This bias might cause the financial impacts of climate change and transitions to be incorrectly priced, and result in an inefficient allocation of capital during the low-carbon transition \citep{doi:10.1002/wcc.678}.

\section{Problem Statement and Related Work}

Recent work has shown first successful computerized analysis of climate-related financial disclosures. For example, the TCFD has conducted an ``AI review,'' using a supervised learning approach that is not further detailed. 
They identified compliance with the TCFD Recommended Disclosures, but did not assess the quality of the disclosed information nor the type of risk \citep{tcfd2019}.   
This approach was refined by \citet{luccioni2020analyzing}, who developed a question answering approach to identify passages in climate disclosures that answer the 14 TCFD recommendations, and make their trained model accessible as a tool for sustainability analysts.
\citet{bingler2021cheap} developed ``ClimateBERT'' to analyze compliance with TCFD recommendations in a variety of corporate reporting globally, and find mostly disclosure of non-material TCFD categories. 
\citet{kolbel2020does} were able to identify an increase in disclosure of transition risks in 10-K reports that outpaced those of physical risks, based on their measure of climate disclosure using a fine-tuned BERT model. 
Finally, \citet{sautner2020firm} use a rule-based approach for identifying CRO-related language in corporate conference calls. They use machine learning (ML) for expanding their set of keywords, which was also proposed by \citet{luccioni2019using} for analyzing climate-related disclosure.

All of these studies have analyzed the number of mentions of climate-related disclosures, however, the quality and materiality of the disclosures remains largely unclear. Analyzing the types of reported risks is a step into this direction, allowing potential investors to better judge the materiality of reported risks. We expand on \citet{kolbel2020does} by introducing more fine-grained risk categories and detect them in free text such as European annual reports (instead of 10-Ks). While most previous work has taken the approach to classify at the sentence level, we observe that more context is needed for disclosing risks, and we classify at the paragraph-level.

To carry out our project, we create a novel dataset based on a refined labeling scheme to distinguish different types of climate-related risks (Section \ref{sec:data}). We then train different classification algorithms to identify and categorize paragraphs in free-text annual reports that disclose such climate-related risks (Section \ref{sec:methods}). 
Finally, we apply the model to analyze climate-related disclosure in annual reports of 337 European firms over the past decade (Section \ref{sec:app}).


\section{Data}\label{sec:data}

We created our own labeled dataset for the task of classifying paragraphs according to disclosure of climate risks.\footnote{Available upon request.} 
We built a corpus of annual reports from the 50 largest publicly traded companies (STOXX Europe 50) and more than half of the European firms in the STOXX Europe 600 index for the last 20 years (where available), which we obtained from the companies' investors relation websites and Refinitiv Eikon \citep{refinitv2021}.
We then parsed the PDF files using the Apache Tika package~\footnote{\texttt{\url{https://tika.apache.org/}}} and split the documents on each page into paragraphs using a rule-based approach (regex). 

The paragraphs were annotated by student assistants familiar with climate policy, who were trained by the authors and followed our code book (see Appendix). The five risk categories include two types of physical risks and three types of transitions risks. ``Acute'' and ``chronic'' physical risks denote those from increases in extreme weather events and those that develop slowly like changes in precipitation patterns, respectively. Transition risks include those related to the potential introduction or strengthening of climate policies (``policy \& legal''), to changing market and technological environments (``tech.~\& market''), and to the reputation of corporations or products (``reputation'').

For the test and validation datasets, we labeled 120 {\it STOXX Europe 50} reports in their entirety. We sampled stratified by years and industries in order to avoid bias in the dataset and later evaluate the model performance across industries and time. To reduce the number of pages to screen, we pre-selected those pages that included at least one match with an extensive list of relevant keywords (see Appendix), and their neighboring pages. On average, a report consisted of 34 relevant pages with 16 paragraphs each. All paragraphs on those selected pages were then annotated with the five categories allowing multiple labels per paragraph. Paragraphs without risk disclosure on relevant pages were considered ``negative examples'', and perceived edge cases were labeled as ``hard negatives.'' We randomly split the dataset in test and validation data, and ensured that each contains a separate set of companies to avoid spill-overs.

The classes are highly imbalanced, which is why we employed a greedier approach for the training dataset focused on covering the variance among positive examples and including only hard negative examples. We extracted relevant pages from annual reports of {\it STOXX Europe 600} companies using a more tailored keyword list than for the test dataset, and then selectively labeled relevant paragraphs.  

The resulting datasets are summarized in \autoref{tab:dataset}. 
To assess the inter-coder reliability, two coders independently labeled 20 reports 
resulting in a Krippendorff's alpha of $\alpha=0.20$ (union) and $\alpha=0.69$ (intersection) for 5 classes. 


\begin{table}[t]
\caption{Number of labeled paragraphs in dataset (some paragraphs have several labels)}
\label{tab:dataset}
\vskip 0.15in
\begin{center}
\begin{small}
\begin{sc}
\begin{tabular}{lcccr}
\toprule
{} &  Train &  Val &  Test \\
\midrule
\multicolumn{1}{l}{\textit{Physical risks}}\\
\hspace{3mm}\bfseries  Acute               &       133 &  15 & 28 \\
\hspace{3mm}\bfseries  Chronic             &        54 &  5 & 19 \\
\multicolumn{1}{l}{\textit{Transition risks}}\\
\hspace{3mm}\bfseries  Policy \& Legal      &      43 & 40 & 60 \\
\hspace{3mm}\bfseries  Tech.~\& Market &        37 &     17 & 21 \\
\hspace{3mm}\bfseries  Reputation          &        23 & 14 & 14 \\
\midrule
\multicolumn{1}{l}{\bfseries  Unique pos.~paragraphs} &       205 & 72 & 97 \\
\midrule
\multicolumn{1}{l}{\bfseries  Neg. paragraphs} &     295 &  39'007 & 40'878 \\
\hspace{3mm}of these \textit{hard neg.}       &       295 &  73 & 55 \\
\bottomrule
\end{tabular}
\end{sc}
\end{small}
\end{center}
\vskip -0.1in
\end{table}

\section{Methods}\label{sec:methods}
\subsection{Tasks and Models}
We divide the task of classifying climate risks in three tasks of increasing difficulty: {\it Binary} (classification in ``risk'' or ``no risk''), and multi-label with {\it two} (physical and transition risks) and {\it five classes} (all risk categories). On all of these tasks we evaluate a baseline model, pretrained DistilBERT~\cite{sanh2019distilbert}, and RoBERTa~\cite{liu2019roberta}.

As a baseline model, we selected a support vector machine (SVM)~\cite{cortes1995support} as a one-versus-rest classifier and applied standard preprocessing to the input such as stop-word removal, lemmatization and TF-IDF weighting. 
We addressed class imbalances between negatives and positives with class weights and used Precision-Recall AuC on the validation set for scoring. 

To leverage context-specific word embeddings, we fine-tuned different variants of pretrained BERT-related models \citep{devlin2019bert} such as DistilBERT \citep{sanh2019distilbert} and RoBERTa Large \citep{liu2019roberta} on our training dataset using negative log-likelihood loss with a softmax activation function for binary classification and a binary cross-entropy loss for the multi-label classification tasks. Again, we calculated class weights to address the class-imbalance.
We trained the models for 4 epochs, using early stopping and limited hyperparameter search on the validation set.
We also determined the optimal class probability thresholds by maximizing the F1-score on the validation set. 
Training is estimated to have emitted less than {$ 7.12\ kg \ \rm{CO}_{2}e$} in total.

\subsection{Experiments and Validation}

We evaluate the models on the three tasks defined above as well as in the following settings: (1) \textit{discriminatory} where no negative paragraphs are present, (2) \textit{hard negatives} setting with paragraphs that are edge cases, and (3) \textit{realistic} setting with all negatives from pre-selected pages. We choose the best model for inference based on the F1-score on the validation set for the realistic setting and five risk categories.

RoBERTa achieves the best performance in the 5-class/realistic setting (Table \ref{tab:val}). As RoBERTa is the largest among the models compared, this confirms expectations. Notably, DistillBERT performs slightly better in the 2-class/realistic setting, which could indicate that with more training examples it might be sufficient to rely on a smaller model. Comparing across settings, the realistic case appears the most difficult, 
and the models perform best on the the discriminatory case (without negatives). Remarkably, SVM outperforms RoBERTa in the easiest setting. We also added additional negative training examples in the realistic setting, which did not improve performance.

\subsection{Test Results}

We evaluate the model performance on held-out test data in the realistic setting (Table \ref{tab:test}). 
In general, the model suffers from a relatively low recall of $0.360$, and its precision does not exceed $0.5$. This is explained by the fact that the task of identifying disclosures requires domain expertise and is also rather difficult for humans. For comparison, after refining the coding scheme, we conducted a review of test and validation data, resulting in a precision of $0.59$ for the preliminary coding of {\it binary}, which is lower than what the model achieves on the same task.
We find a large variance in performance across classes, with reputational transition risks being the hardest to identify (F1-score of $0.140$) and acute physical risks the easiest (F1-score of $0.537$). 
We also observe that physical risk classes exhibit a considerably higher precision and lower recall than transition risk classes, which are more balanced. 





\begin{table}[t]
\caption{Validation performance (F1-score macro-avg.)}
\label{tab:val}
\vskip 0.15in
\begin{center}
\begin{small}
\begin{sc}
\begin{tabular}{lccc}
\toprule
Experiment & SVM &  DistilBERT &  RoBERTa \\
\midrule
\multicolumn{1}{l}{\textit{5 Classes}}\\
\hspace{3mm} \bfseries  Realistic              &       0.204 &     0.241 &     \bfseries 0.356\\
\hspace{3mm}  Hard neg.             &        0.457 &    0.431   &     0.528\\
\hspace{3mm} Discriminatory             &        0.599 &      0.558 &     0.596\\
\midrule
\multicolumn{1}{l}{\textit{2 Classes}}\\
\hspace{3mm} Realistic      &       0.351 &     0.497 &     0.446\\
\midrule
\multicolumn{1}{l}{\textit{Binary}}\\
\hspace{3mm} Realistic      &       0.290 &    0.444 &     0.496\\
\bottomrule
\end{tabular}
\end{sc}
\end{small}
\end{center}
\vskip -0.1in
\end{table}

\begin{table}[t]
\caption{Test performance for RoBERTa (best model) in the realistic setting for 5 classes and binary.}
\label{tab:test}
\vskip 0.15in
\begin{center}
\begin{small}
\begin{sc}
\begin{tabular}{lccr}
\toprule
{} &  Precision & Recall & F1 \\
\midrule
\multicolumn{1}{l}{\textit{Physical risks}}\\
\hspace{3mm}\bfseries  Acute               &       0.846 &     0.393 &     0.537\\
\hspace{3mm}\bfseries  Chronic             &        0.833 &      0.263 &     0.400\\
\multicolumn{1}{l}{\textit{Transition risks}}\\
\hspace{3mm}\bfseries  Policy \& Legal      &       0.291 &     0.383 &     0.331\\
\hspace{3mm}\bfseries  Tech. \& Market &        0.400 &     0.476 &     0.435 \\
\hspace{3mm}\bfseries  Reputation          &        0.093 &      0.286 &     0.140\\
\midrule
\multicolumn{1}{l}{\bfseries  Avg.~5 classes} &       0.493 &     0.360 &  0.369\\
\midrule
\multicolumn{1}{l}{\bfseries  Avg.~Binary} &       0.695 &     0.423 &  0.526\\
\bottomrule
\end{tabular}
\end{sc}
\end{small}
\end{center}
\vskip -0.1in
\end{table}

\section{Applying the Model}\label{sec:app}
We determined the number of risk mentions for 4,498 annual reports by all 337 companies in the dataset by performing inference on pages in proximity of a keyword match. Out of 2.7m analyzed paragraphs, 3892 paragraphs were predicted to contain at least one risk (total of 5501 risk mentions).

\begin{figure}[ht]
\centering
\subfigure[2 classes]{\label{acror_cro__years}\includegraphics[width=0.49\columnwidth]{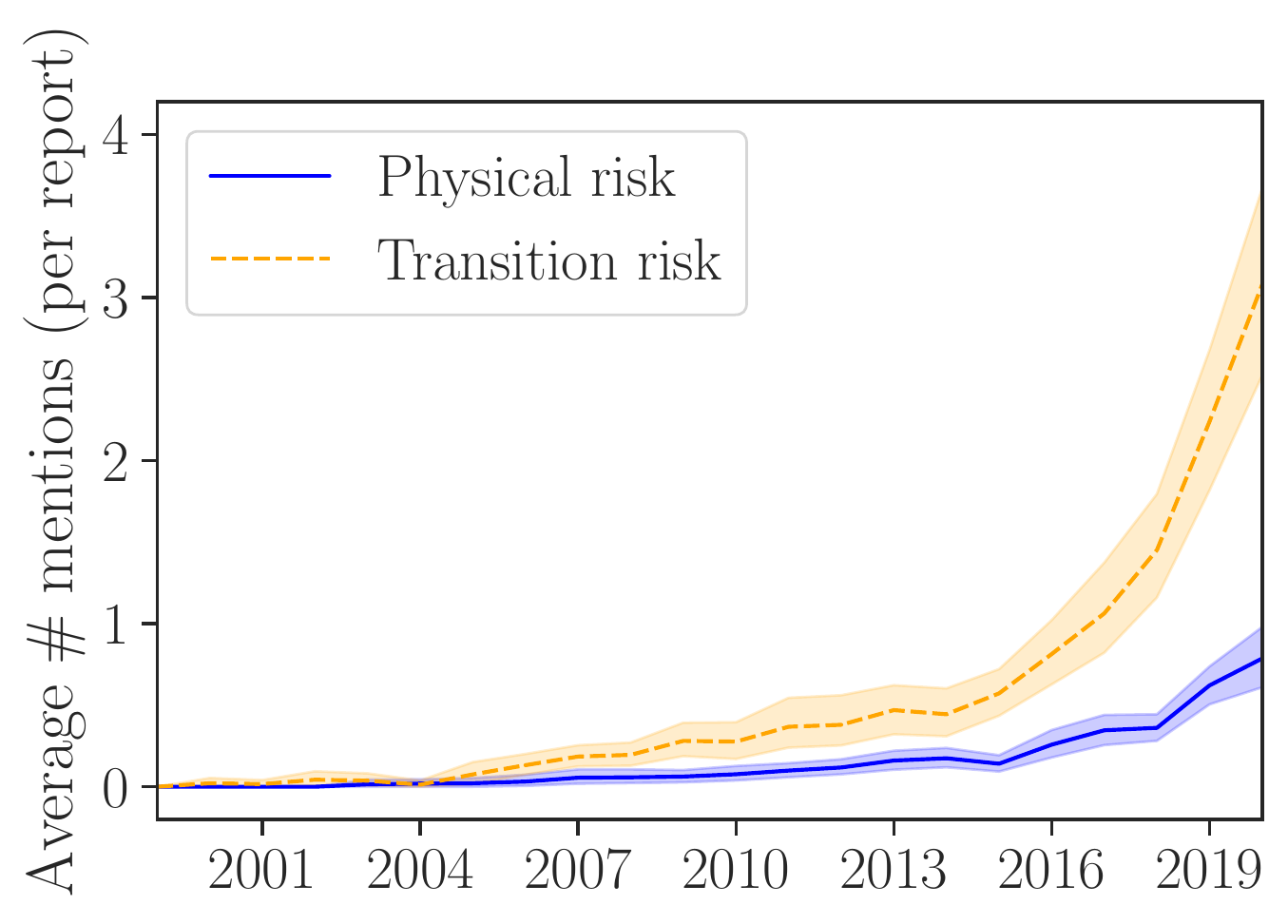}
}
\subfigure[5 classes]{\label{acror_cro_sub_type_years}\includegraphics[width=0.46\columnwidth]{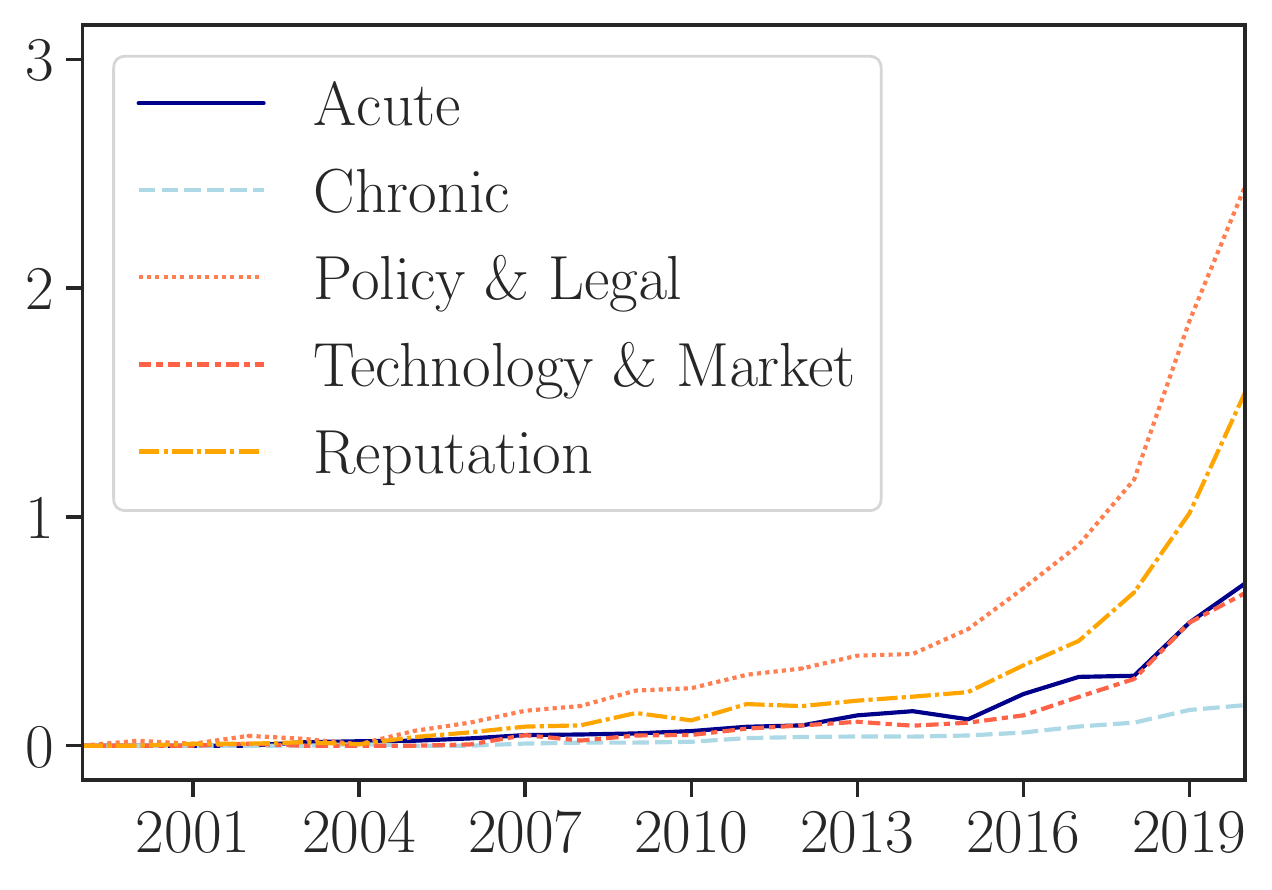}}
\caption{Average mentions per report over time. a) with 2 classes and a 95\%-bootstrapped confidence interval (CI), b) with 5 classes.}
\end{figure}

Figure \ref{acror_cro__years} shows the average number of mentions per report of physical and transition risks over time, which grew slowly until 2015, after which it increased rapidly. This growth is particularly high for transition risks, resulting in about three times as many mentions compared to physical risks in 2019. The analysis of risk subcategories (Figure \ref{acror_cro_sub_type_years}) reveals that the growth was mainly driven by ``policy \& legal'' and ``reputation'' risks. 

\begin{figure}[ht]
\vskip 0.2in
\begin{center}
\centerline{\includegraphics[width=\columnwidth]{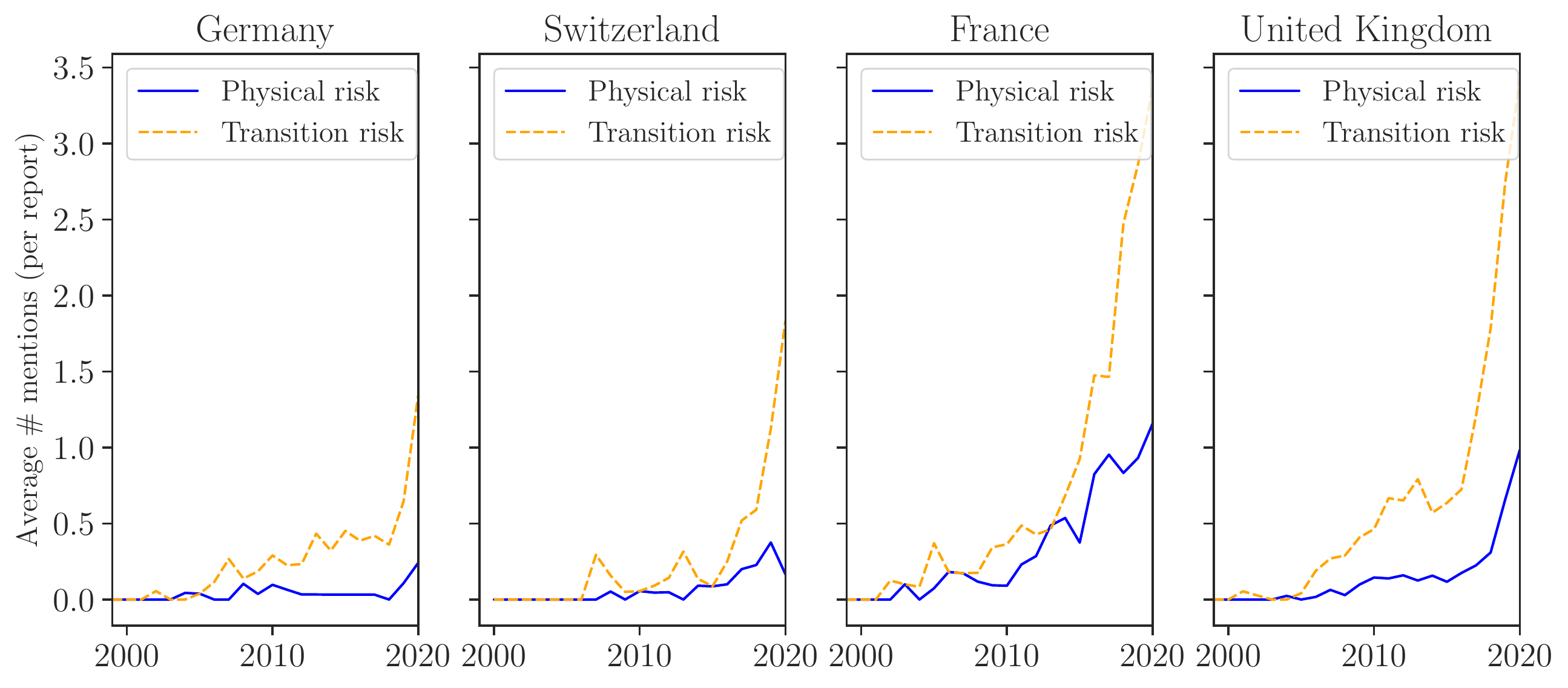}}
\caption{Average number of climate risk mentions per report for selected countries.}
\label{acror_selected_countries}
\end{center}
\vskip -0.2in
\end{figure}

\begin{figure}[ht]
\vskip 0.2in
\begin{center}
\centerline{\includegraphics[width=\columnwidth]{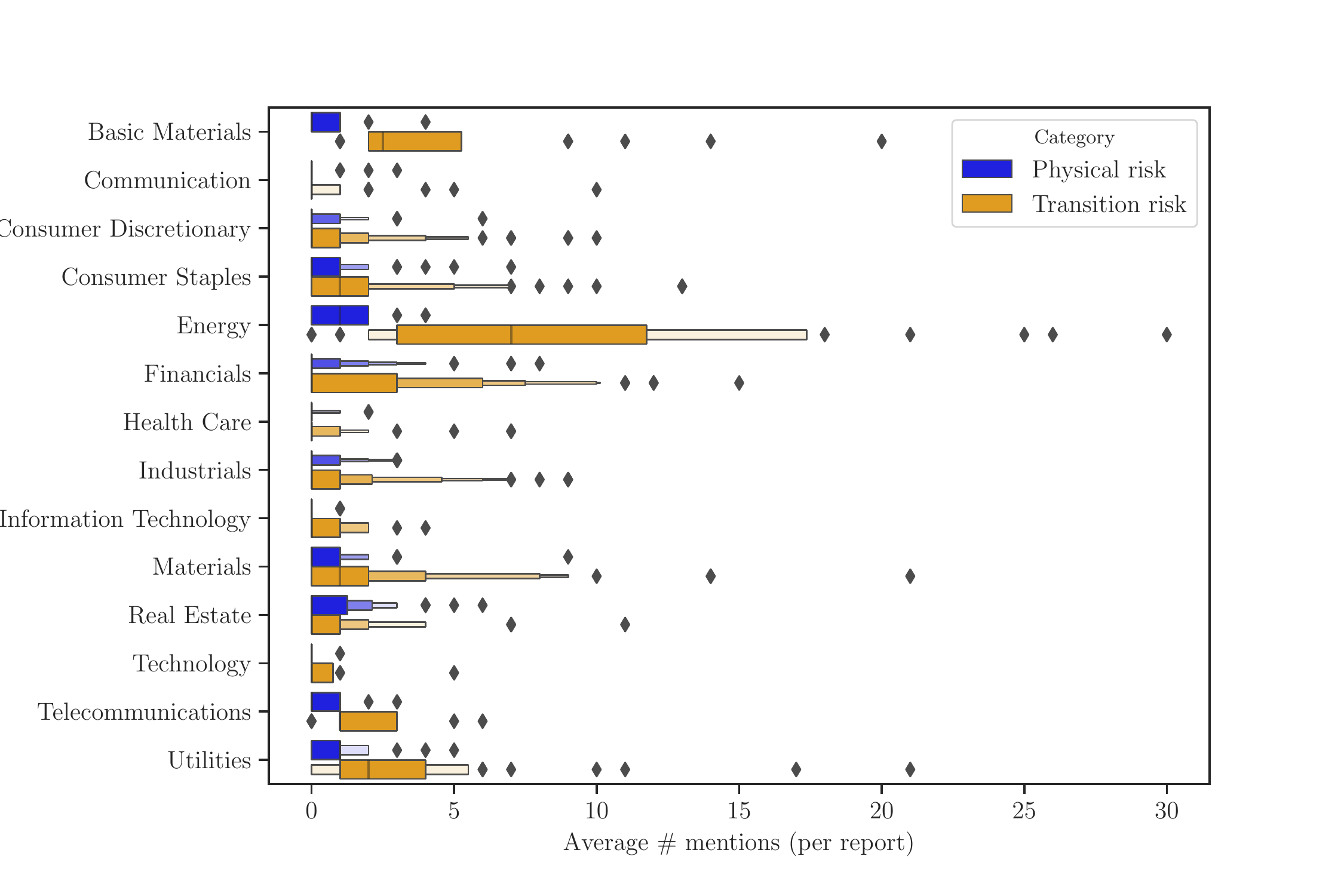}}
\caption{Distribution of the average number of climate risk mentions per report over the time frame 2015-2020 by industry.}
\label{acror_industry_dist}
\end{center}
\vskip -0.2in
\end{figure}

Given different regulatory environments, we compare the dynamics in four countries in Figure \ref{acror_selected_countries}. Companies in France, which has a disclosure mandate, and the United Kingdom saw a marked rise in both transition and physical risk reporting since 2015, while Germany and Switzerland exhibited a lower (but still clearly visible) growth during the same period. Comparing different industries (Figure \ref{acror_industry_dist}), we find that especially the energy, basic materials, and utilities industries disclose transition risks. These are sectors with high emission intensities, which are particularly affected by climate policies. For physical climate risk, no clear industry pattern is visible in our data.

\section{Discussion and Conclusion}

In the present article, we developed an approach to automatically identify climate risk disclosures in corporate annual reports, and used it to analyze disclosure of 337 European companies over 20 years. 
We find that the number of risk mentions (especially of transition risks) started to rise sharply around 2015. It appears likely that public policies played a role for this development, as numerous policies to encourage or mandate climate risk reporting have been enacted in Europe since 2015 \citep{steffen2021comparative}.
To assess whether specific policies indeed caused the development, however, requires further research. Potential empirical designs to that end include difference-in-differences approaches, or models with country- and industry-fixed effects. Our approach is well suited to the deliver the dependent variable for such analysis.

Next steps for refining our analysis will focus on appropriately quantifying the uncertainty of model predictions, and working to reduce it further by exploring a hierarchical classification approach, and adding more training data.
The analysis can also be expanded to a broader set of company types and communication channels beyond annual reports. 

Finally, it should be kept in mind that improved transparency on climate risks should not automatically be expected to change investor behavior in a meaningful way \citep{ameli2020climate}. More research is needed to understand how capital is (re)allocated based on better climate risk disclosures; in this context our approach can be useful to deliver the explanatory variable for such future analyses. Ultimately, both the effectiveness of policies to trigger climate risk disclosures, and the effectiveness of such disclosures to change investment behavior, are required for financial investments to help achieve the targets of the Paris Agreement.






\section*{Acknowledgements}
The project has received funding from the European Union’s Horizon2020 research and innovation programme, European Research Council (ERC) (grant agreement No 948220, project GREENFIN).


\bibliography{main.bib}
\bibliographystyle{icml2021}

\end{document}


\onecolumn
\icmltitle{Appendix: Climate risk disclosure in annual reports}

\section{Coding scheme}

\textbf{Physical Risk}: Risks from the physical impacts of climate-change. Note: Include also ``indirect'' physical risks, i.e.~when climate change is not mentioned directly but reference to one of physical effects from climate change (second order effects). 
\begin{itemize}
\item Acute: Risks from Increase severity and/or frequency of extreme weather events, cyclones, floods, heat waves.
\item Chronic: Risks from changes in precipitation patterns (droughts), rising mean temperatures, rising sea levels.
\end{itemize}

\textbf{Transition Risk}: Description of risks from the transition to a lower-carbon economy.

\begin{itemize}
    \item Policy \& Legal: Risks related to the potential introduction or strengthening of climate policies, such as carbon tax, emission reporting policy changes, regulation on products and services, litigation risk (i.e. climate-related lawsuits).
    \item Technology \& Market: Risks related to changing market environments because of climate change, such as changing customer behavior, uncertainty in market prices, increase of cost in raw material \& natural resources. Also includes technology risks from climate change-related obsolescence of existing products \& services, unsuccessful investments in new technologies, costs to transition in to lower emission technology.  
    \item Reputation: Risks to the reputation of the corporation or its products/services because of climate change-related matters, such as changing customer preferences, stigmatization of sector (including by investors), hiring risk, increased stakeholder pressure.
\end{itemize}

\section{Keywords}

To filter relevant pages in annual reports, we have used the following sets of keywords (comma-separated) using lemmatized comparison.

\underline{General terms:} 
climate change,
global warming,
climate risk,
greenhouse effect,
sustainable energy,
renewable,
carbon,
co2,
co2e,
ghg,
greenhouse,
climate mitigation,
paris agreement,
kyoto protocol,
ipcc,
climate adaptation,
changing climate.

\underline{Specific terms pertaining to transition risks:}
emission regulation,
emission standard,
emission reduction,
emission trading,
cap and trade,
oil price,
energy price,
fossil fuel,
energy legislation,
environmental legislation,
climate legislation,

\underline{Specific terms pertaining to physical risks:}
natural hazard,
windstorm,
floods,
floodings,
drought,
global temperature,
temperature rise,
extreme weather,
sea level,
disaster,
extreme event,
storm,
hurricane,
biodiversity,
rainfall,
rain,
monsoon,
catastrophic event,
climate feedback,
climate impacts,
climate variability.

\section{Inference}

Table \ref{tab:SI} shows summary statistics for the model inference results.

\begin{table}[t]
\caption{Number of risks identified in annual reports.}
\label{tab:SI}
\vskip 0.15in
\begin{center}
\begin{small}
\begin{sc}
\begin{tabular}{llrrrr}
\toprule &                &            &          & Average \# mentions per report &          &     \\
\bfseries                 & \bfseries             & \bfseries  Coverage & \bfseries                  Mean & \bfseries  St. Dev. & \bfseries  Max \\
{} & {} &          &                      &          &     \\
\midrule
\textbf{Physical risk} & \textbf{Acute} &      492 &                 1.51 &     0.96 &   8 \\
                & \textbf{Chronic} &      189 &                 1.12 &     0.34 &   3 \\
\textbf{Transition risk} & \textbf{Policy \& Legal} &     1002 &                 2.52 &     2.77 &  26 \\
                & \textbf{Technology \& Market} &      390 &                 1.60 &     1.15 &  10 \\
                & \textbf{Reputation} &      711 &                 1.96 &     1.71 &  14 \\
\bottomrule
\end{tabular}
\end{sc}
\end{small}
\end{center}
\vskip -0.1in
\end{table}